\documentclass[aps,prl,twocolumn,preprintnumbers,superscriptaddress,nofootinbib,10pt]{revtex4-1}

\usepackage{amssymb}
\usepackage{amsmath}
\usepackage{amsfonts}
\usepackage{graphicx}
\usepackage{subfigure}
\usepackage{color}
\usepackage{dcolumn}
\usepackage{hyperref}
\newcommand{\be}{\begin{equation}}
\newcommand{\ee}{\end{equation}}
\newcommand{\bea}{\begin{eqnarray}}
\newcommand{\eea}{\end{eqnarray}}

\begin{document}

\title{Real-Time dynamics and phase separation in a holographic first order phase transition}

\author{Romuald A. Janik}
\email{romuald@th.if.uj.edu.pl}
\affiliation{Institute of Physics, Jagiellonian University, Lojasiewicza 11, 30-348  Krakow, Poland}
\author{Jakub Jankowski}
\email{jjankowski@fuw.edu.pl}
\affiliation{Faculty of Physics, University of Warsaw, ulica Pasteura 5, 02-093 Warsaw, Poland}
\author{Hesam Soltanpanahi}
\email{hsoltan@ipm.ir}
\affiliation{School of Physics, Institute for Research in Fundamental Sciences (IPM), P.O.Box 19395-5531, Teheran 19538-33511, Iran}


\begin{abstract}
We study the fully nonlinear time evolution of a holographic system possessing a first
order phase transition. The initial state is chosen in the spinodal
region of the phase diagram, and includes an inhomogeneous perturbation
in one of the field theory directions. The final state of the time evolution
shows a clear phase separation in the form of domain formation.
The results indicate the existence of a very rich class of inhomogeneous black hole
solutions. 
\end{abstract}

\maketitle

\noindent {\bf Introduction.}
The concept of a phase transition is 
inherently an equilibrium one and it is a
theoretical challenge to formulate a framework
in which it can be quantitatively studied
in cases involving real time dynamics.
Such a framework is offered by gauge-gravity
duality put forward 
about twenty years ago \cite{Maldacena:1997re}.
Gauge-gravity 
duality, also referred to as holography,
is immanently related to field theory systems at strong coupling
and thus it is very useful to study nonperturbative physics
especially as within this framework one can work
directly in Minkowski signature and study real time dynamics.
Recently, it has been demonstrated that well 
known holographic $1^{\rm st}$ order phase transitions \cite{Gubser:2008ny}
are accompanied by an unstable spinodal region 
\cite{Janik:2015iry,Janik:2016btb} in accordance with  
standard predictions \cite{Chomaz:2003dz}.
The analysis was based on linear response theory, however
the true power of the dual gravitational approach is 
the possibility of investigating fully nonlinear dynamic 
evolution of the system. Simultaneously with the present work,
this particular direction
has been recently undertaken in Ref. \cite{Attems:2017ezz}
where spinodal region was studied in the holographic model
of a phase transition. The final state of the time evolution
was an inhomogeneous configuration approached at late
times within the hydrodynamic approximation.
A different set-up of a homogeneous evolution 
has been undertaken in Ref. \cite{Gursoy:2016ggq}.
Inhomogeneous, static configurations appearing in the context
of a holographic first order phase transition
were also recently studied in Ref. \cite{Dias:2017uyv}.

A natural question which arises is whether the final state 
will exhibit domains of the two coexisting phases with the same
values of the free energy.
The main result of the present letter is to demonstrate for the first time
that in the case of a three-dimensional nonconformal system with 
a holographic dual such a phase separation will arise dynamically
through a real time evolution from a perturbation in the spinoidal
region.
The respective energy densities of the two components
of the final state are very close to the
corresponding energy densities determined at the critical 
temperature. This implies that the system undergoes a
dynamical transition during which different regions 
of space become occupied with different phases of matter.

\begin{figure}[h!]
	\begin{center}
	\includegraphics[height=.23\textheight]{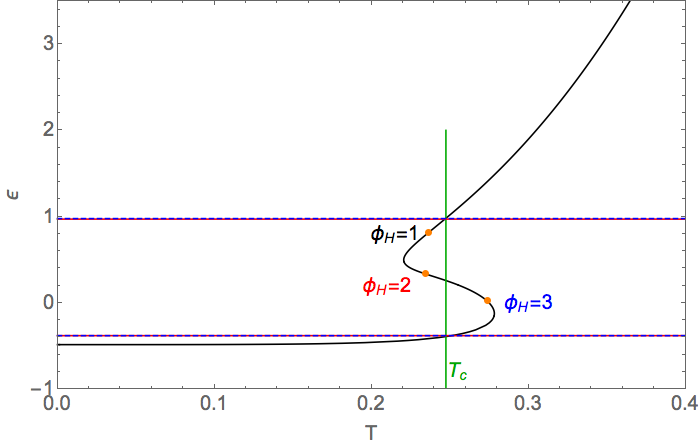}
	\includegraphics[height=.23\textheight]{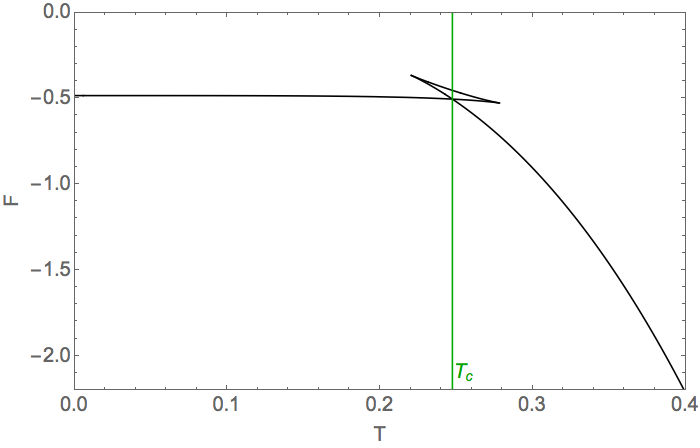}
	\caption{Upper panel: The temperature dependence of the holographic energy density.
	 Vertical green line represents the transition temperature.
	 Orange points represent sample chosen initial configurations for time evolution.
	 Horizontal lines show the domain energies in the final state 
	(solid - cosine perturbation, dashed - Gauss perturbation).
	 Lower panel: free energy as a function of temperature.}
	\label{fig:eT}
	\end{center}
\end{figure}

\noindent {\bf The holographic framework.}
The holographic model we use is a bottom-up construction
containing Einstein gravity coupled to a real, self-interacting
scalar field. The action takes the standard form
\begin{equation}
S=\frac{1}{2\kappa_4^2}\int d^4x \sqrt{-g}  \left[ R-\frac{1}{2}\, \left( \partial \phi \right)^2 - V(\phi) \, \right] + S_{\rm GH} + S_{\rm ct}~,
\label{eq:Action}
\end{equation}
where is $\kappa_4$ related to the four dimensional
Newton's constant $\kappa_4=\sqrt{8\pi G_4}$, and the self interaction
potential
$V(\phi)$ is given below. We include the 
boundary terms in the form of Gibbons-Hawking $S_{\rm GH}$ \cite{Gibbons:1976ue}
and holographic counterterms
$S_{\rm ct}$ \cite{Skenderis:2002wp,Elvang:2016tzz} contributions.
We chose to work in 3+1-dimensional bulk spacetime, which is
dual to 2+1-dimensional field theory,
for the absence of conformal anomaly
in odd dimensions. This, in turn, makes the
expansions near the conformal boundary free from logarithms,
which allows for the usage of Chebyshev spectral methods
for numerical integration.
The $V(\phi)$ potential
is constructed so that the dual field theory
undergoes an equilibrium first order phase transition.
The specific choice that we use is
\begin{equation}
V(\phi)=-6 \cosh \left(\frac{\phi}{\sqrt{3}}\right)+b_4\phi^4~,
\label{eq:V}
\end{equation} 
where $b_4=-0.2$. This functional form is dual to a relevant deformation
of the boundary conformal field theory with an operator of conformal dimension
$\Delta=2$. When $b_4=0$ the potential is that of $\mathcal{N}=2$
supergravity in $D=4$ after dimensional reduction from $D=11$ \cite{Duff:1999gh}.
The physical scale, breaking  conformal invariance, 
is set by the source of the operator, and is chosen to be $\Lambda=1$.
The equilibrium structure of this model is described in terms of 
dual black hole geometries characterized by specifying the 
horizon value of the scalar field, i.e., $\phi(z=1)=\phi_H$.
The entropy and the free energy of the system are, in turn, given
by the Bekenstein-Hawking formula, and the on-shell value of the
action (\ref{eq:Action}) respectively.
The corresponding thermodynamics 
reveals the appearance of a first order phase transition
between different branches of black hole geometries as determined
by the difference of free energies. The order of the transition
is established by a discontinuity of the first derivative
of the free energy of the system. 
This effect is illustrated in the lower panel 
of Fig. \ref{fig:eT}.
The value of the critical temperature 
is $T_c\simeq0.246$ (in $\Lambda=1$ units). This transition is of 
similar nature as the Hawking-Page transition in the case of anti-de Sitter (AdS) space
\cite{Hawking:1982dh,Witten:1998zw} (an important difference, however, is that
in the current setup all phases are of the black hole type). 
The equation of state (EOS) is displayed
in the upper panel of Fig. \ref{fig:eT} as a temperature dependence of the energy 
density. This EOS is similar to the five-dimensional
case studied in \cite{Janik:2015iry,Janik:2016btb}
and the detailed analysis of the linearized dynamics proved, 
in accordance with the general lore, the existence of a spinodal
region separating stable configurations \cite{Janik:2015iry,Janik:2016btb}.  


\noindent {\bf Time dependent configurations.}
To study the time evolution of the system we adopt the 
following metric ansatz in the Eddington-Finkelstein (EF)
coordinates
\begin{equation}
ds^2=-A\,dt^2-\frac{2\,dt\,dz}{z^2}-2\,B\, dt\,dx + S^2\,\left( G\,dx^2+G^{-1}\, dy^2 \right)~,
\label{eq:Metric}
\end{equation}
where all functions are $x,t,z$ dependent.
We can take the initial state to lie in the spinodal 
region of the phase diagram and add an $x$-dependent
perturbation to the $S$ function. By the proper 
choice of the perturbing function
\begin{equation}
\delta S(t,x,z) = S_0\,z^2\,(1-z)^3\,\cos\left(k x\right)~, 
\label{eq:Sperturbation}
\end{equation} 
we can turn on a particular unstable mode or
add a mixture of all the modes
\begin{equation}
\delta S(t,x,z) = S_0\,z^2\,(1-z)^3\,\exp\left[-w_0 \cos\left(\tilde{k} x\right)^2\right]~,    
\label{eq:GaussPert}
\end{equation}
with different widths.
By solving the  time dependent Einstein-matter 
equations of motion with proper $AdS$ boundary conditions at $z=0$
\footnote{In the EF coordinates $A\sim1/z^2+O(1)$, $S\sim1/z+O(1)$, 
$G\sim O(z)$, $B\sim O(z)$ for $z\rightarrow0$.} we determine
the non-linear evolution of the system.
Using the procedure of holographic renormalization we then read-off
the relevant observables like the energy density of the the boundary theory from
subleading terms in the near-boundary expansion \cite{Skenderis:2002wp,Elvang:2016tzz}. 

The system is essentially studied in the microcanonical ensemble as
the total energy density of the system is fixed throughout the evolution.
The gravitational formulation of the problem 
is now given by a coupled set of non-linear 
partial differential equations. We solve that
problem numerically using the characteristic formulation
of General Relativity \cite{Chesler:2013lia}
along with spectral methods \cite{Grandclement:2007sb}.
In the relevant spatial direction we use 
periodic boundary conditions with spectral
Fourier discretization. The remaining spatial direction
is uncompactified.


\noindent {\bf Results.}
Since for this system the energy density in equilibrium 
uniquely determines the temperature,
we may deduce that the final state of evolution starting from the unstable spinoidal branch 
necessarily has to be inhomogeneous. The physical expectation that the final state will 
consist of well separated phases at the phase transition temperature $T=T_c$ would manifest 
itself in the existence of spatial domains characterized by very flat energy density,
the values of which should coincide with the energy densities of the two stable phases
at $T=T_c$. The fact that such a configuration is attained dynamically even when starting
from points on the spinoidal branch with temperatures differing from $T_c$ is far from trivial. This is the main
result of the present work.

To illustrate the effect of the appearance of different
phases during the time evolution we run the simulations
for a number of initial configurations,
covering a representative region of interest.
Some of the configurations are marked with orange
dots in Fig. \ref{fig:eT}. As it was explained in the 
previous section we use two different shapes of 
perturbing function given in Eq. (\ref{eq:Sperturbation}) and (\ref{eq:GaussPert})
with different values for parameters.
Particularly transparent results appear for the value
of the momentum equal to $k=1/6$ and $\tilde{k}=1/12$, with $w_0=10$, and we have chosen to 
present those in this letter.
The amplitudes of the perturbations are in the range $S_0=0.1- 0.5$.

The first point of interest is a large black hole configuration with temperature below $T_c$, 
but still on the stable branch
e.g. with $\phi_H=1$.
The linear analysis shows no instability of that
configuration. However, one could still expect a non-linear
instability due to overcooling. In our simulations we found, however, no evidence
of that within the considered framework.

\begin{figure}
	\begin{center}
	\includegraphics[height=.23\textheight]{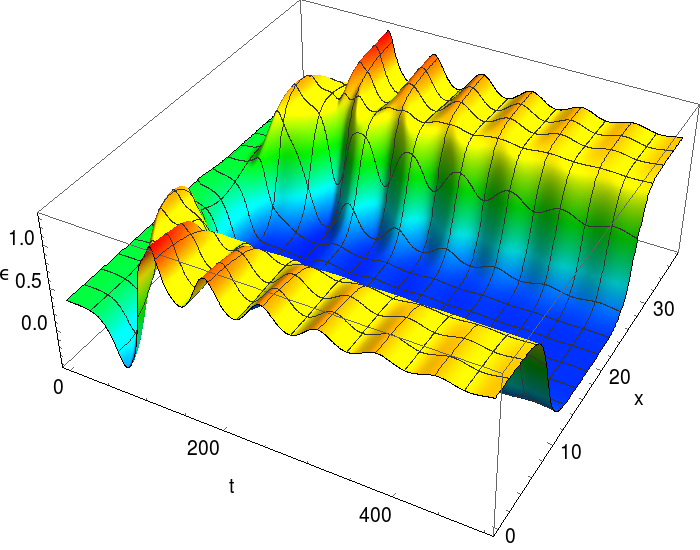}
  \includegraphics[height=.23\textheight]{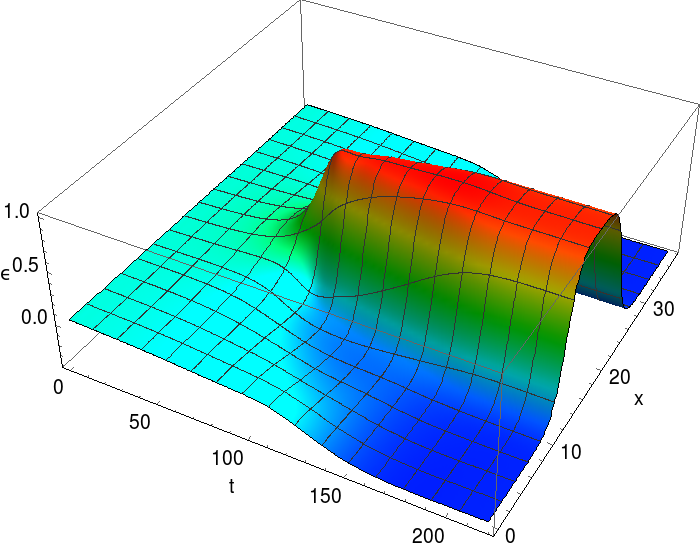} 
	\caption{The energy density as a function of time for the initial
	configuration in spinodal region: $\phi_H=2$ Gaussian perturbation (upper panel), $\phi_H=3$ cosine perturbation (lower panel).
	}
	\label{fig:EnergyP2}
	\end{center}
\end{figure}

The second considered point, with $\phi_H=2$, is placed
deep in the unstable region. The temporal and spatial
dependence of the energy density is shown in 
the upper panel of Fig. \ref{fig:EnergyP2}.
Initially small Gaussian perturbation grows with time, and after
around $400$ units of simulation time starts settling
down to an inhomogeneous final state. The maximum and the
minimum energy of this state, marked as horizontal solid 
lines in Fig. \ref{fig:eT}, approach within less than $1\%$ the energy
densities determined by the transition temperature.
Pronounced flat regions of constant energy density are
apparent at late stages of the evolution.
It is clearly visible that in the final state
different parts of the system are occupied by the
different phases, joined by a domain wall.


The third considered configuration, with $\phi_H=3$,
lies close to the end of the spinodal region.
The time dependence of the energy density in this case
is displayed in the lower panel of Fig. \ref{fig:EnergyP2}.
The perturbation added is a single cosine mode. In this case, after
about $150$ units of simulation time, the configuration settles
down to an inhomogeneous final state. The maximum and 
the minimum energy densities of this final state are marked with
horizontal, dashed lines in Fig. \ref{fig:eT}.
Similarly to the previous configuration  the extrema of energy density
reach the corresponding densities determined at the 
transition temperature $T=T_c$. However, due to the fact that the initial energy
density is smaller than the energy density of a configuration
with $\phi_H=2$, we observe a smaller region of the high-energy
phase in the final state.


\begin{figure}
	\begin{center}
	\includegraphics[height=.23\textheight]{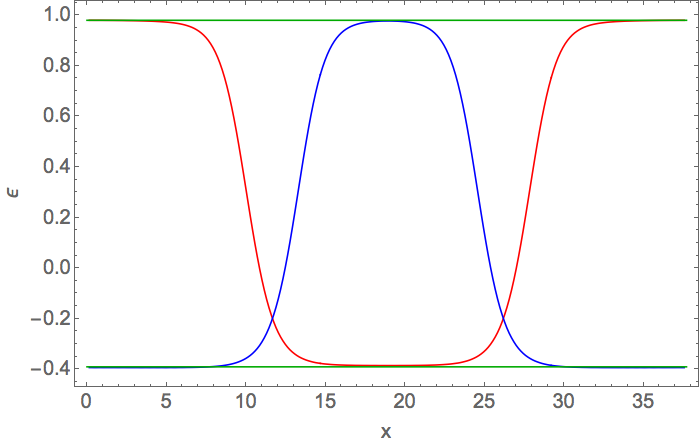}
	\caption{%
	The final state energy density for different initial configurations: $\phi_H=2$ (red) and $\phi_H=3$ (blue).
	The horizontal green lines represent the energy densities at the critical temperature in isotropic solutions.}
	\label{fig:EnergyFinal}
	\end{center}
\end{figure}

\begin{figure}
	\begin{center}
	\includegraphics[height=.19\textheight]{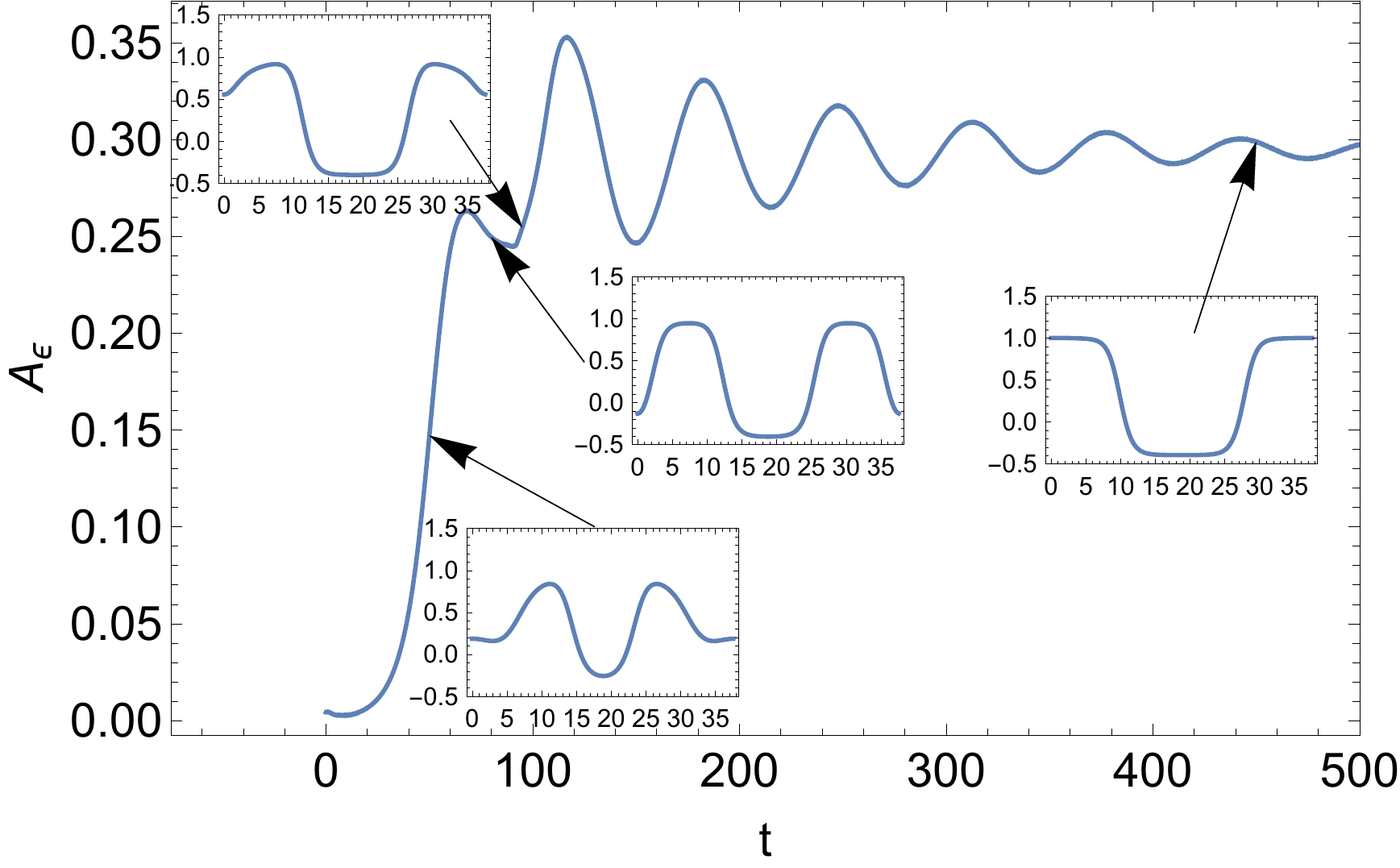}\\
	\medskip
	\includegraphics[height=.19\textheight]{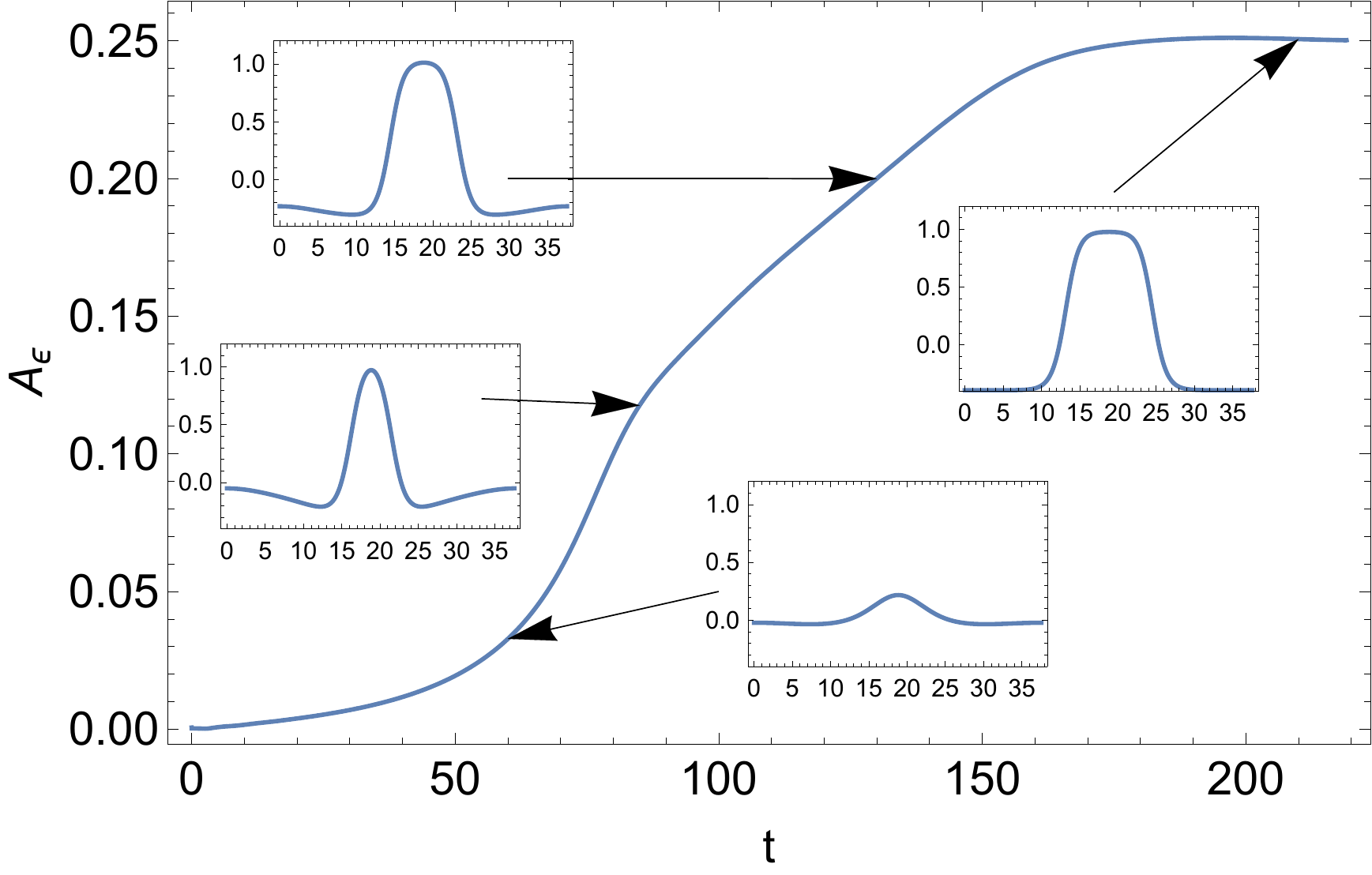}
	\caption{The time evolution of the observable $A_\epsilon$ defined in Eq.~\ref{eq:Aepsilon}.
	 Upper panel: configuration starting from $\phi_H=2$. Lower panel: configuration starting from $\phi_H=3$.
	 The insets show profiles of the energy density at given instants of time.
	 See discussion in text.
	}
	\label{fig:dynamics}
	\end{center}
\end{figure}

To summarize the above results we display the final sate energy density
as a function of $x$ in Fig. \ref{fig:EnergyFinal} for
both unstable initial configurations. The Hawking temperature of 
the final state geometry is constant and equal to the critical temperature $T_c$.
From the field theory perspective this is a clear demonstration of the
coexistence phenomenon, where inhomogeneous state has constant temperature.
Different regions of space are occupied by different 
phases connected by domain walls.
Despite the fact that in both cases the final state is rather 
universal the time evolution is substantially different. 
The great novelty of our approach is that details of 
dynamical formation of domains of different phases can
be studied quantitatively. In order to do so it is convenient to introduce
the following observable
\begin{equation}
A_\epsilon = \frac{1}{12\pi}\int_{\epsilon>\epsilon_0}\epsilon(t,x)dx~,
\label{eq:Aepsilon}
\end{equation} 
where $\epsilon_0$ is the mean energy of the system at $t=t_0$. The above quantity essentially 
measures the amount of energy \emph{above} the mean energy stored in the system at $t=t_0$.
As is clearly seen in Fig. \ref{fig:dynamics} in both considered cases the details
of dynamics are different\footnote{These differences stem from the different forms of the
perturbation employed by us at $\phi_H=2$ and $\phi_H=3$.}. For the configuration with $\phi_H=2$ the
initial perturbation develops into two bubbles which subsequently move away from the center 
and then violently coalesce and merge into one
final domain. Because of that, the final state is approached with large, damped oscillations.
In contrast, the configuration with $\phi_H=3$ displays less violent evolution. We can identify
three stages of evolution in that case. First is an exponential growth of the instability,
that takes place roughly
until the maximum energy reaches the equilibrium energy density. 
The second stage displays linear increase of the bubbles width with 
fixed height to form an extended region. The system finally saturates in the 
third stage with small oscillations
for late times. In both cases the complicated dynamics is a consequence of non-linear nature
of dual Einstein's equations, and would be extremely difficult to study using conventional field theory techniques.
In future work we intend to analyze in more detail both of these phenomena: the collision of fully formed
domains of equilibrium phases as well as the details of the dynamics of bubble growth.


\noindent {\bf Discussion.}
In this letter we demonstrated for the first time 
the existence of a phase separation effect at the
transition temperature of a first order phase transition 
in the context of holographic models. The chosen set-up was a bottom-up
holographic construction designed to exhibit an equilibrium
first order phase transition.


\begin{figure}
	\begin{center}
	\includegraphics[height=.23\textheight]{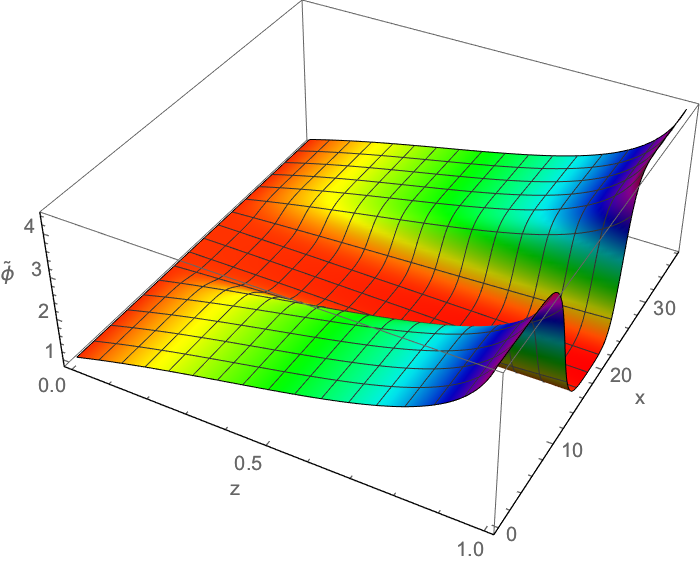}
	
	\caption{The final configuration of the scalar field $\phi(x,z)/z$ extending to the
	inhomogeneous horizon at $z=1$ for the solution starting from $\phi_H=3$.
	}
	\label{fig:bulkphi}
	\end{center}
\end{figure}


The full nonlinear time evolution of a perturbed
configuration with a spinodal instability ends in an inhomogeneous 
state composed of domains of different stable phases as determined
at the transition temperature fixed by the equality of free energies.
The dual gravitational configuration is a black hole with
an inhomogeneous horizon. However in contrast to the results of~\cite{Attems:2017ezz},
the geometries obtained here correspond to domains of specific thermodynamic phases with very
mild spatial dependence separated by relatively narrow domain walls. 
On the gravitational side this means that the bulk geometry consists of
two distinct types of black holes, characterized e.g. by the value of the
scalar field on the horizon smoothly connected by interpolating domain walls 
(see Fig.~\ref{fig:bulkphi} for an example configuration of the scalar field in the bulk).
Despite the inhomogeneity, the Hawking temperature is constant on the horizon. 
From the field theory interpretation we expect to have an immense moduli
space of geometries which correspond to different configurations of 
phase domains coming from different seed perturbations.
Indeed, we found also solutions with multiple potential domains
which, however, were of size of the order of the domain wall width
and thus were more similar to the solutions of~\cite{Attems:2017ezz}.
It is clear that with an appropriately large transverse spatial extent
of the simulation one can explicitly construct solutions with multiple domains.
We intend to investigate this in the future. 
 
We have also demonstrated that
large black holes with temperatures below the critical 
value are stable against the perturbations that we tried.
This indicates that in the holographic description, we may 
naturally expect an overcooled phase which only starts
to nucleate once it has dynamically moved into the spinoidal branch.

There are numerous directions for further research. Apart from studying
the space of domain geometries mentioned earlier, it would be very
interesting to investigate in detail the various temporal regimes
which can be seen in Fig. \ref{fig:dynamics}
and study the dynamics of the phase domains.
Naturally it would be good to investigate extensions to other dimensions as well as ultimately
relax any symmetry assumptions.
An additional bonus of the present setup is the appearance 
configurations breaking translation invariance
without specifying explicit inhomogeneous sources.
In other words, the final state solutions spontaneously break
translational invariance. This opens a possibility
of applications in the context of condensed matter
physics \cite{Horowitz:2012ky,Donos:2013eha}.


\noindent{\bf Acknowledgments.} 
We would like to thank Z. Bajnok for extensive discussions and collaboration in the initial stage
of the project.
RJ  was supported by the NCN grant 2012/06/A/ST2/00396. JJ was supported by the
NCN grant No. UMO-2016/23/D/ST2/03125.
JJ and HS would like to thank theory division at CERN, Wigner institute and Jagiellonian University where
a part of this research was carried out. HS would like to thank to Witwatersrand university where part of this research was done.
RJ would like to thank the Galileo Galilei Institute for Theoretical Physics for
hospitality and the INFN for partial support during the completion of this work.



\begin{thebibliography}{99}


\bibitem{Maldacena:1997re}
  J.~M.~Maldacena,
  The Large N limit of superconformal field theories and supergravity,
  Adv.\ Theor.\ Math.\ Phys.\  {\bf 2} 231 (1998). 
	
	
\bibitem{Gubser:2008ny} 
  S.~S.~Gubser and A.~Nellore,
  Mimicking the QCD equation of state with a dual black hole,
  Phys.\ Rev.\ D {\bf 78}, 086007 (2008).
	
\bibitem{Janik:2015iry} 
  R.~A.~Janik, J.~Jankowski and H.~Soltanpanahi,
  Nonequilibrium Dynamics and Phase Transitions in Holographic Models,
  Phys.\ Rev.\ Lett.\  {\bf 117}, no. 9, 091603 (2016).
	
\bibitem{Janik:2016btb} 
  R.~A.~Janik, J.~Jankowski and H.~Soltanpanahi,
  Quasinormal modes and the phase structure of strongly coupled matter,
  JHEP {\bf 1606}, 047 (2016).
	
\bibitem{Chomaz:2003dz} 
  P.~Chomaz, M.~Colonna and J.~Randrup,
  Nuclear spinodal fragmentation,
  Phys.\ Rept.\  {\bf 389}, 263 (2004).
	
\bibitem{Attems:2017ezz} 
  M.~Attems, Y.~Bea, J.~Casalderrey-Solana, D.~Mateos, M.~Triana and M.~Zilhao,
  Phase Transitions, Inhomogeneous Horizons and Second-Order Hydrodynamics,
	JHEP {\bf 1706}, 129 (2017).
		
\bibitem{Gursoy:2016ggq} 
  U.~G{\"u}rsoy, A.~Jansen and W.~van der Schee,
  New dynamical instability in asymptotically anti-de Sitter spacetime,
  Phys.\ Rev.\ D {\bf 94}, no. 6, 061901 (2016).

\bibitem{Dias:2017uyv} 
  O.~J.~C.~Dias, J.~E.~Santos and B.~Way,
  Localised and nonuniform thermal states of super-Yang-Mills on a circle,
  JHEP {\bf 1706}, 029 (2017).

		
\bibitem{Gibbons:1976ue} 
  G.~W.~Gibbons and S.~W.~Hawking,
  Action Integrals and Partition Functions in Quantum Gravity,
  Phys.\ Rev.\ D {\bf 15}, 2752 (1977).

\bibitem{Skenderis:2002wp} 
  K.~Skenderis,
  Lecture notes on holographic renormalization,
  Class.\ Quant.\ Grav.\  {\bf 19}, 5849 (2002).

\bibitem{Elvang:2016tzz} 
  H.~Elvang and M.~Hadjiantonis,
  A Practical Approach to the Hamilton-Jacobi Formulation of Holographic Renormalization,
  JHEP {\bf 1606}, 046 (2016).
	
	
\bibitem{Duff:1999gh} 
  M.~J.~Duff and J.~T.~Liu,
  Anti-de Sitter black holes in gauged N = 8 supergravity,
  Nucl.\ Phys.\ B {\bf 554}, 237 (1999).

\bibitem{Hawking:1982dh} 
  S.~W.~Hawking and D.~N.~Page,
  Thermodynamics of Black Holes in anti-De Sitter Space,
  Commun.\ Math.\ Phys.\  {\bf 87}, 577 (1983).
	
\bibitem{Witten:1998zw} 
  E.~Witten,
  Anti-de Sitter space, thermal phase transition, and confinement in gauge theories,
  Adv.\ Theor.\ Math.\ Phys.\  {\bf 2}, 505 (1998).
	
	
\bibitem{Chesler:2013lia} 
  P.~M.~Chesler and L.~G.~Yaffe,
  Numerical solution of gravitational dynamics in asymptotically anti-de Sitter spacetimes,
  JHEP {\bf 1407}, 086 (2014).


\bibitem{Grandclement:2007sb} 
P.~Grandclement and J.~Novak,
Spectral methods for numerical relativity,
Living Rev.\ Rel.\  {\bf 12}, 1 (2009).

\bibitem{Horowitz:2012ky} 
  G.~T.~Horowitz, J.~E.~Santos and D.~Tong,
  Optical Conductivity with Holographic Lattices,
  JHEP {\bf 1207}, 168 (2012).

\bibitem{Donos:2013eha}
  A.~Donos and J.~P.~Gauntlett,
  Holographic Q-lattices,
  JHEP {\bf 1404} (2014) 040.

\end{thebibliography}
\end{document}